\documentclass{appolb}
\usepackage{epsfig}
\def\be{\begin{eqnarray}}
\def\ee{\end{eqnarray}}
\def\prt{\partial}
\def\lsim{\mathrel{\rlap{
\lower4pt\hbox{\hskip-3pt$\sim$}}
\raise1pt\hbox{$<$}}}     %less than approx. symbol
\def\gsim{\mathrel{\rlap{
\lower4pt\hbox{\hskip-3pt$\sim$}}
\raise1pt\hbox{$>$}}}     %greater than or approx. symbol
\begin{document}
\title{Chiral magnetic effect and electromagnetic field evolution %
\thanks{Presented at the 28$^{\rm th}$ Max Born Symposium and HIC for FAIR 
Workshop ``Three Days on Quarkyonic Island'', Wroclaw, May 19-21, 2011}%
}
\author{V.~D.~Toneev and V.~Voronyuk
\address{Joint Institute for Nuclear Research, Dubna, Russia \\ 
Frankfurt Institute for Advanced Studies, Frankfurt, Germany}
}
\maketitle
\begin{abstract}
The energy dependence of the observable two-particle correlator in search 
for local strong parity violation in Au+Au collisions is estimated within a 
simple phenomenological model. 
The model reproduces available RHIC data but predicts that at LHC the
chiral magnetic effect (CME)  will be about 20 times weaker than at RHIC,
contrary to the first  LHC measurements. 
In the lower energy range  this effect should vanish sharply at an energy 
somewhere above the top SPS one in agreement
with the preliminary results of the Beam Energy Scan  program. 
To elucidate CME background effects a transport HSD model
including magnetic field evolution is put forward and electromagnetic dynamics
at RHIC energies is investigated. 
It is observed that the  electromagnetic field included into the hadronic model
does not influence on observables due to mutual compensation of effects of 
electric and magnetic fields.
\end{abstract}

\section{Introduction}

The existence of nontrivial topological
configurations  in QCD vacuum is a fundamental property of the gauge
theory though until now there is no direct experimental evidence for
topological effects.
Transitions between different topological states occur with the
change of the topological number $n_w$ characterizing these states
and induce anomalous processes like
local violation of the ${\cal P}$ and ${\cal CP}$ symmetry.
 The interplay  of these topological
configurations with (chiral) quarks results in the local imbalance
of chirality. Such chiral asymmetry coupled to a strong magnetic
field, created by colliding nuclei perpendicularly to the reaction
plane, induces a current of electric charge along the direction of
a magnetic field which leads to a separation of oppositely charged
particles with respect to the reaction plane. Thus, as was argued
in Refs.~\cite{Kharzeev:2004ey,KZ07,KMcLW07,FKW08,KW09} the
topological effects in QCD may be observed in heavy ion collisions
directly in the presence of very intense external magnetic
fields due to the ``Chiral Magnetic Effect'' (CME) as a
manifestation of spontaneous violation of the ${\cal CP}$
symmetry. 
It was shown that the electromagnetic field of the required
strength can be created in relativistic heavy-ion
collisions~\cite{KMcLW07,SIT09}. First experimental evidence for
the CME identified via the observed charge separation effect with
respect to the reaction plane was presented by the STAR
Collaboration at RHIC~\cite{Vo09}. Recently preliminary data 
 below the nominal RHIC energy (Beam Energy Scan
program)~\cite{BES11} and at the LHC energy~\cite{ALICE11} 
have been obtained.

In this paper we shortly remain our analysis of the first STAR data~\cite{TV10}
and compare results with new finding. In respect to the CME, we also make
a dynamical estimate of the CME background based on the nonequalibrium
Hadron-String-Dynamics (HSD) microscopical transport approach \cite{HSD}
supplemented by the treatment of electromagnetic field evolution.

\section{CME in a simple model}
We remind here our phenomenological model used in the CME analysis
in Ref.~\cite{TV10}.

For one-dimensional random walk in the topological number space
the topological charge (winding number) $n_w$ generated
during the time $\tau_B$, when the magnetic field is present, may be
estimated as
\begin{equation}
\label{def1}
n_w \equiv \sqrt{Q_s^2}=\sqrt{ \Gamma_{S} \cdot V \cdot \tau_B} \sim
\sqrt{\frac{dN_{\rm hadrons}}{dy}} \cdot \sqrt {Q_s \ \tau_B}~,
\end{equation}
where $\Gamma_S$ is the sphaleron transition rate which in
 weak and strong
coupling $\Gamma_S \sim  T^4$ with different coefficients. The
initial temperature $T_0$ of the produced matter at time $\tau
\simeq 1/Q_s$ is proportional to the saturation momentum $Q_s$,
$T_0 = c\ Q_s$. In the last semiequality of (\ref{def1}) the
expansion time and the corresponding time dependence of the
temperature are neglected. Since sizable  sphaleron transitions
occur only in the deconfined phase, the time $\tau_B$ in
Eq.~(\ref{def1}) is really the smallest lifetime between the
strong magnetic field $\tilde{\tau}_B$ one and the lifetime of
deconfined matter $\tau_{\epsilon}$: $\tau_B = {\rm
min}\{\tilde{\tau}_B, \tau_{\epsilon}\} $.

The measured electric charge particle asymmetry  is associated with the 
averaged correlator $a$ by the following relation~\cite{Vol05}:
\begin{eqnarray}
\label{cos}
\langle \cos (\psi_\alpha+\psi_\beta-2\Psi_{RP}) \rangle =
 \langle \cos (\psi_\alpha+\psi_\beta-2\Psi_c) \rangle / v_{2,c}=
v_{1,\alpha} v_{1,\beta} - a_\alpha a_\beta~,
\end{eqnarray}
where $\Psi_{RP}$ is the azimuthal angle of the reaction plane defined by
the beam axis and the line joining the centers of colliding nuclei.
Averaging in (\ref{cos})  is carried out over the whole event ensemble.
The second equality in (\ref{cos}) corresponds to azimuthal measurements
with respect to particle of type $c$ extracted from three-body correlation
analysis~\cite{Vol05}, $v_1$ and $v_2$ are the directed and elliptic
flow parameters, respectively.  According  to Ref.~\cite{Kharzeev:2004ey}
an average correlator $a=\sqrt{a_\alpha a_\beta}$ is related to the
topological charge, $n_w$, as
\begin{equation}
\label{rel}
a \sim \frac{n_w}{dN_{\rm hadrons}/dy} \sim
\frac{\sqrt {Q_s \tau_B}}{\sqrt{dN_{\rm hadrons}/dy}}
\sim \sqrt{\frac{\tau_B}{Q_s}} \sim (\sqrt{s_{NN}})^{-1/16}
\cdot \sqrt{\tau_B},
\end{equation}
where absorption and rescattering in dense matter are neglected
for the same and opposite charge correlations. In the last
equality we assumed that $Q_s^2 \sim s_{NN}^{1/8}\sim dN_{\rm
hadrons}/dy$~\cite{KN01}.

\begin{figure}[h]
\centering\includegraphics[angle=-90,width=0.6\textwidth]{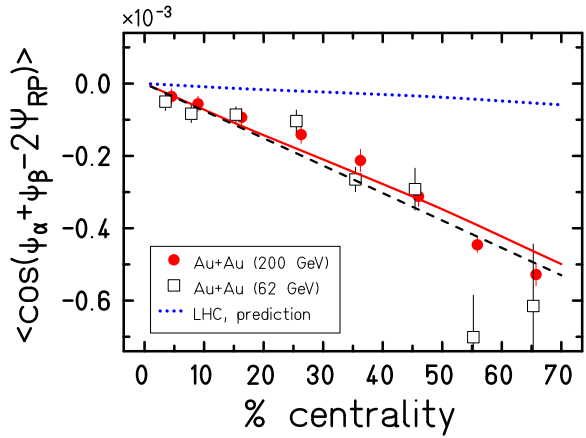}
\caption{Centrality dependence of the CME. Experimental points for
Au$+$Au  collisions are from~\cite{Vo09}. The dotted line is our
prediction for Au+Au collisions at the 5.5 TeV energy.
 \label{CME_cu}}
\end{figure}

As follows from  Eq.~(\ref{rel}), the
 $\cal{CP}$ violation effect can be quantified by the correlator
\begin{equation}
\label{res3}
a^2  =  K_{Au} \  (\sqrt{s_{NN}})^{-1/8} \cdot \tau_B~.
\end{equation}
The normalization constant $K_{Au}$ can be tuned  at the reference
energy $\sqrt{s_{NN}}=$200 GeV from the inverse relation
(\ref{rel}) and experimental value $a_{exp}$ at this energy for
$b=$10 fm
\begin{equation}
\label{res4}
K_{Au}=\frac{ a^2_{exp} \cdot (200)^{1/8} } { \tau_B(200)}~.
\end{equation}
The only quantity needed is the lifetime $\tau_B$ defined as the time
during which the
magnetic field is above the critical value $B_{crit}$ needed to support a
fermion Landau level on the domain wall $eB_{crit} = 2 \pi/ S_d$,
where $S_d$ is the domain wall area. Since the size of the domain
wall is not reliably known, it is hard to pin down the number, but
it should be of the order of $m_\pi^2$. Thus, we have to treat
$B_{crit}$ as a free parameter which defines the lifetime from the calculated
$B(t)$ distribution~\cite{TV10}. As to the impact parameter distribution,
 the CME is assumed  to be roughly linear in $b/R$ Ref.~\cite{KMcLW07}.
Taking this as a hypothesis we evaluate the centrality dependence
of the CME fitting this line to points $b=10$ fm (or centrality
$(40-50)\%$) to be estimated in our model and $b=0$ where the CME
is zero. The results are presented in Fig.~\ref{CME_cu} for Au+Au
collisions.
As is seen, the calculated lines quite reasonably reproduce the
measured points of azimuthal asymmetry of same-charge particles
for Au$+$Au collisions at $\sqrt{s_{NN}}=$200 and 62 GeV. The
chosen critical field value $eB_{crit}=0.7\ m_\pi^2$ results in
absence of the CME above the top SPS energy because  the critical
magnetic field practically coincides with the maximal field at
this bombarding energy. This finding is in a qualitative agreement
with very recent preliminary STAR results that the difference
between same-charge and opposite charge correlations is decreasing
with decreasing beam energy what takes place at $\sqrt{s_{NN}}<$40
GeV~\cite{BES11}.

In the model considered, the CME at the energy 5.5 TeV is expected to be
less by a factor of about 20  as compared to those at the RHIC 
energy~\cite{BES11}.
Note that at the LHC energy we applied a simplified semi-analytical
model~\cite{SIT09} for magnetic field creation. Thus, our LHC estimate
 should be considered as an upper limit for the CME. Recent preliminary
 LHC PbPb data~\cite{ALICE11} show
a remarkable agreement in both the magnitude and the behavior with the results
reported by STAR in Au-Au collisions at $\sqrt{s_{NN}} =$200 GeV.
Preliminary kinetic calculations at the LHC energy 2.76 TeV
increase the maximal magnetic field by a factor about three but it does not
influence essentially on the relaxation time.

\section{Kinetic consideration}

The discussed CME signal, the electric charge
asymmetry with respect to the reaction plane,  can originate not
only from the spontaneous local {\cal CP} violation but also be
simulated by other possible effects. In this respect it is important
to consider the CME background. We shall do that considering a full
evolution of  nucleus-nucleus collisions in terms of the HSD transport
model \cite{HSD} but including formation of electromagnetic field as
well as its evolution and impact on particle propagation.

Generalized on-shell transport equations for $N$ strongly interacting particles
in the presence of electromagnetic fields can be written as~\cite{VTB11}
 \be \label{kinEq}
 \left\{ \frac{\prt}{\prt t}+
 \left(e \vec{E} +{ (\frac{e}{c})\ \vec{v}\times
 \vec B} \right)\nabla_{\vec{p}} \
 \right\} \ f(\vec{r},\vec{p},t)=I_{coll}(f,f_1,...f_{N-1})~,
\ee
 which are supplemented by the wave equation for the magnetic field whose
solution in the semi-classical approximation for point-like moving
charges is reduced to the retarded Li\'enard-Wiechert
potential~\cite{SIT09}. The quasiparticle propagation in the electromagnetic field
is calculated according to $d\vec{p}/dt=
e\vec{E}+ (e/c)\ \vec{v}\times \vec{B}.$

\begin{figure}[thb]
\centering
\includegraphics[height=5.5cm,width=0.48\textwidth]{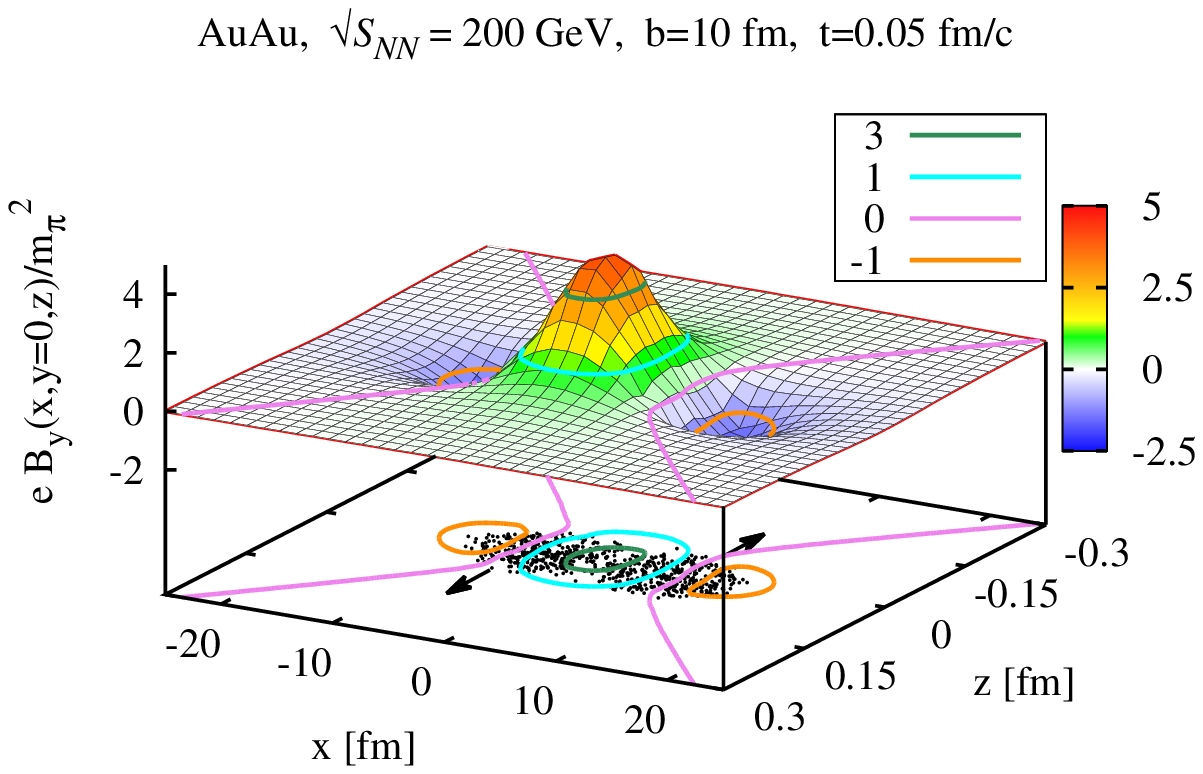}
\includegraphics[height=5.5cm,width=0.48\textwidth]{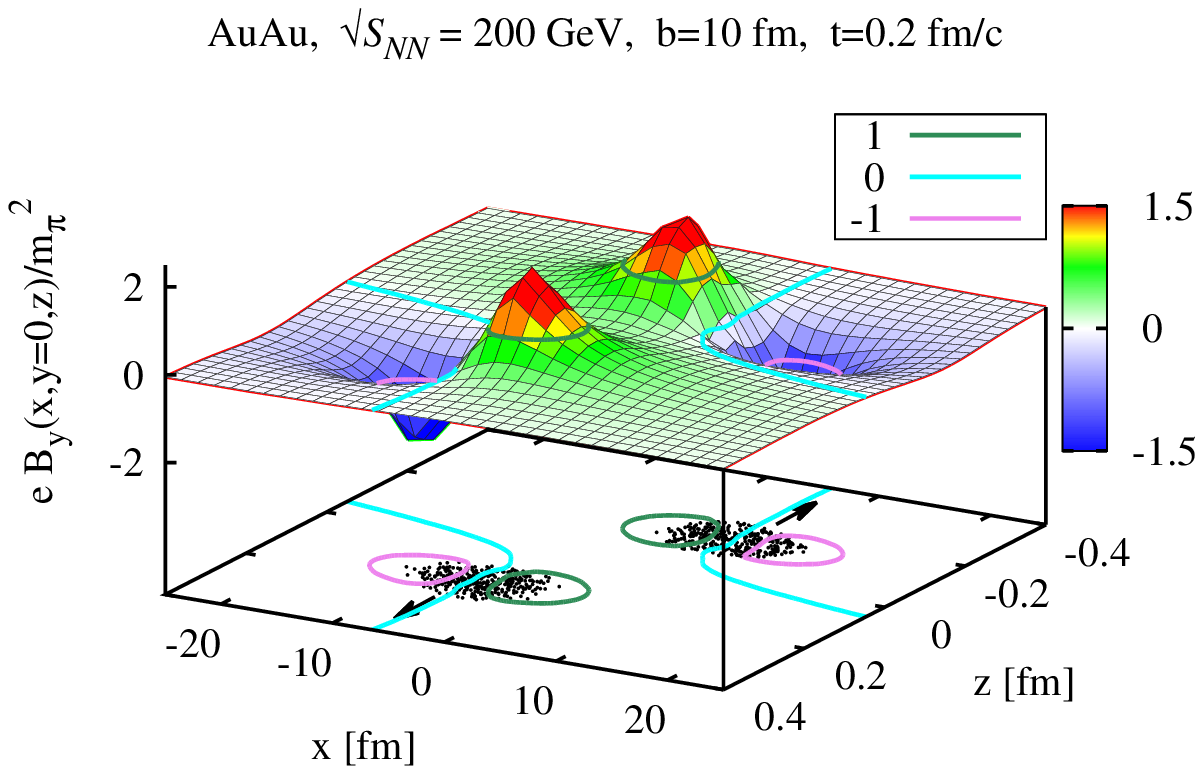}
\caption{Distribution of the magnetic field strength $eB_y$ in the
$y=0$ plane at $t=$0.05 and 0.2 fm/c for Au+Au collisions at
$\sqrt{s_{NN}}=$200 and $b=$10 \ fm. The collision geometry is
projected on $x-z$ plane by points corresponding to a particular
spectator position. Curves (and their projections) are levels of
the constant $eB_y$.
  \label{By} }
\end{figure}

\begin{figure}[thb]
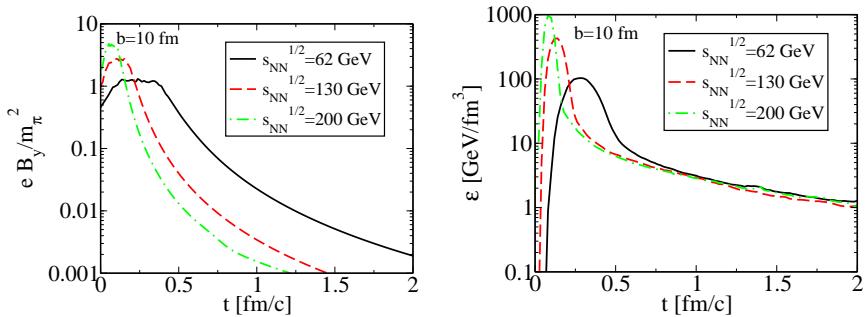

\centering
\includegraphics[width=0.43\textwidth]{B_semi_log_RHIC.eps}
\hspace{0.3cm}\includegraphics[width=0.43\textwidth]{E_semi_log_RHIC.eps}
\caption{Time evolution of the magnetic field strength $eB_y$ (left)
and energy density (right) at the central point $x=$0, $y=0$ and $z=0$
 fm/c for Au+Au collisions at various RHIC energies.
  \label{ByEt} }
\end{figure}

 In  a nuclear collision, the magnetic field will be a
superposition of solenoidal fields from different moving charges.
It is illustrated in Fig.~\ref{By}. Its first panel  is taken at a
quite early compression stage with $t=$0.05 fm/c close to the 
maximal overlapping where the magnetic field
here is maximal. The overlapping strongly interacting participant 
region has an ``almond''-like shape. The nuclear region
outside this almond corresponds to spectator matter which is the
dominant source of the electromagnetic field at the very beginning
of the nuclear collision. Note that in the HSD code the particles
are subdivided into  target and projectile spectators and
participants not geometrically but dynamically: spectators are
nucleons which suffered yet no collision. The next time moment corresponding
to expansion stage is illustrated in
Fig.~\ref{By} and in more detail studied in Ref.~\cite{VTB11}.
The time evolution of the magnetic field and energy density at the
center of the almond region $eB_y(x=0,y=0,z)$ for Au+Au collisions
for the colliding energy $\sqrt{s_{NN}}=$200 GeV and the impact
parameter $b=$10 fm   is shown in Fig.~\ref{ByEt}. It is seen that
the largest values of $eB_y\sim 5 m_\pi^2$ are reached  for a very 
short time. Note that this is an
extremely high magnetic field, since $m_\pi^2 \approx 10^{18} {\rm
gauss}$. Then, the system expands  and the magnetic field
decreases. It is of interest to note that in our  transport model,
the spectator contribution to the magnetic field is practically
vanishing at the center for $t\approx $1 fm/c. In subsequent times the magnetic
field $eB_y$ is formed essentially due to produced participants
with roughly equal number of negative and positive charges which
approximately compensate each other~\cite{VTB11}.
The evolution of the  energy density of created particles is presented 
in the right panel of Fig.~\ref{ByEt}. Here the maximal energy
density (in the center of the colliding system) is $\varepsilon >$
50 GeV/$fm^3$ at the moment of maximal overlap of the nuclei. When
the system expands, it takes a sausage-like shape   
and the energy density drops fast. But even at the
time $t\sim 0.5$ fm/c  the local
energy density  is seen to be above an effective threshold of a
quark-gluon phase transition $\varepsilon\gsim$1 GeV/$fm^3$.
As shown in~\cite{VTB11} the location of the maximum
energy density  correlates with that for the magnetic field.

\begin{figure}[thb]
\includegraphics[height=5.50truecm,width=6.0truecm] {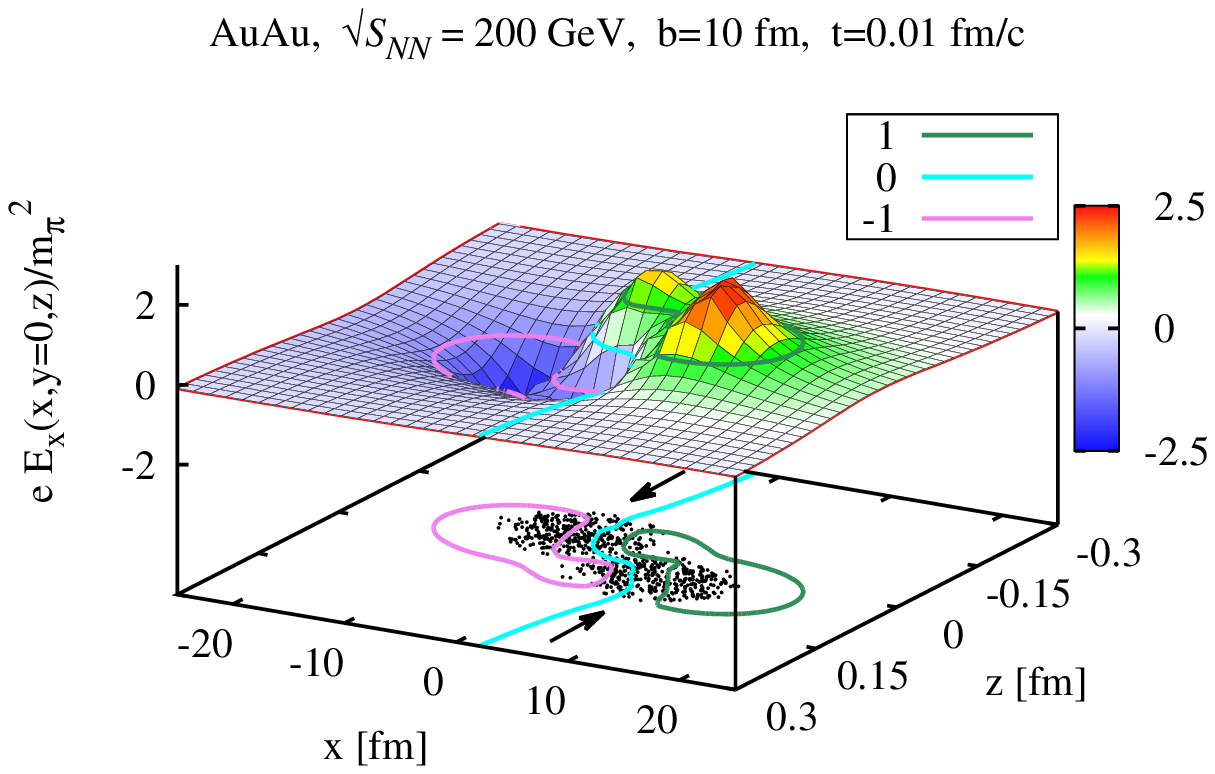}
\hspace{0.1cm}
\includegraphics[height=5.50truecm,width=6.0truecm] {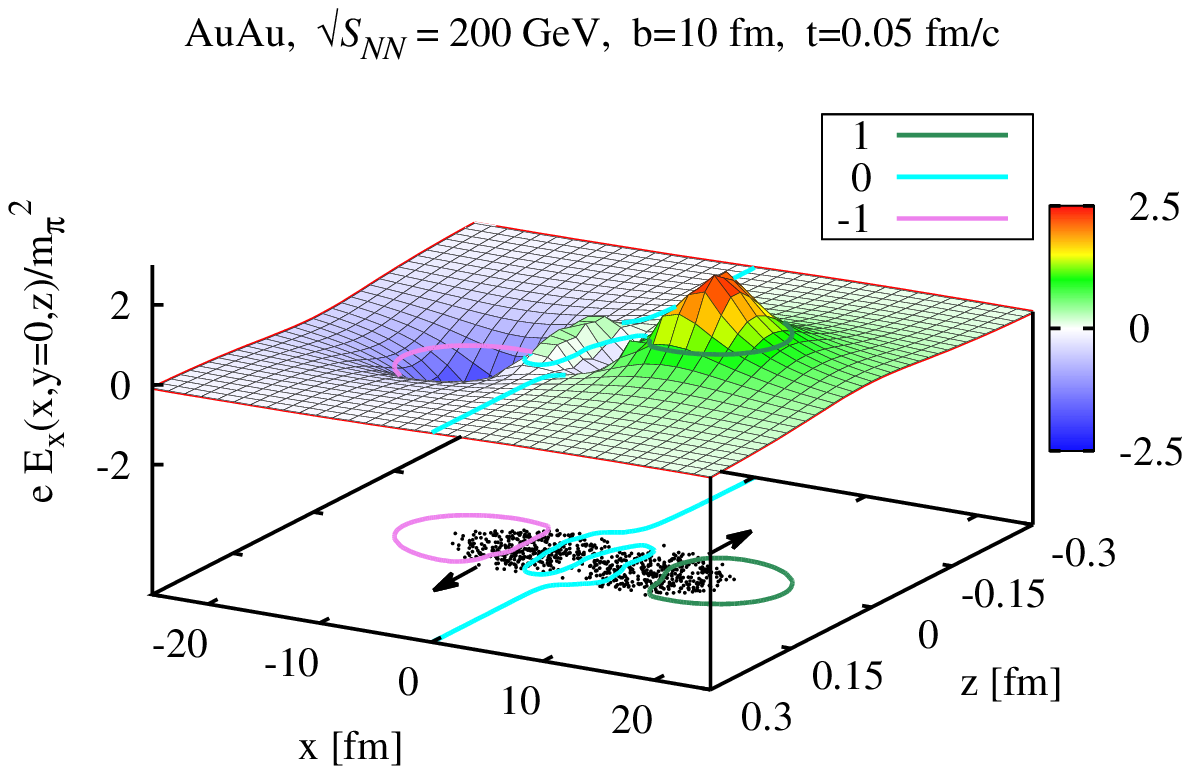}
 \caption{(Color online) Evolution of the $x$- and $y$-components of
the electric field at incoming and maximal overlap in
Au+Au($\sqrt{s}$= 200 GeV) collisions at the impact parameter $b=$10
fm. The $eE_x=$const levels and spectator points are  shown in
the projection on the ($x-z$) plane.
 }
\label{Ex}
\end{figure}

The background electric field, being orthogonal to the magnetic
one, is directed along the $x$ axis.  The evolution of the $eE_x$
field for peripheral ($b=$10 fm) collisions of Au+Au at the top
RHIC energy is presented in Fig.~\ref{Ex}. Similar to the case of
the magnetic field, the $eE_x(x,y=0,z)$ distribution is also
inhomogeneous and closely correlates with geometry while the field
strength looks ``hedgehog'' shaped. When the two nuclei collide,
the electric fields in the overlap region significantly compensate
each other, and the electric field $\vec{E}$ in the target and
projectile spectator parts have opposite signs.
As a result, the locations of the maximum/minimum
are not in the central point of the overlap region - as they are
for the magnetic field - but shifted slightly outside. The maximum
of the electric field can be quite large. All these features are
seen explicitly in Fig.~\ref{Ex} where the temporal evolutions of
$eE_x(x,0,z)$ and $eE_y(x,0,z)$ are given. Due to destructive interference or
the ``hedgehog'' effect, the electric field in the central part of
the overlap region ($x\approx 0$ fm) is consistent with zero apart
from a short period just before reaching maximal overlap. Note
that the electric field at the central point is negligible for 
$t\gsim 0.15$ fm/c.

\section{Observables and electric charge separation}

The HSD model quite successfully describes many observables in a
large range of the collision energy. Here we investigate to what
extent the electromagnetic field - incorporated in the HSD
approach -  will affect some observables. We shall limit ourselves
to Au+Au collisions at $\sqrt{s_{NN}}= 200$ GeV and impact
parameter $b=10$ fm. Here we calculate the whole nuclear
interaction including the decays of resonances at least up to
times of 50 fm/c.

\begin{figure}[th]
\includegraphics[height=4.5truecm,width=6.0truecm] {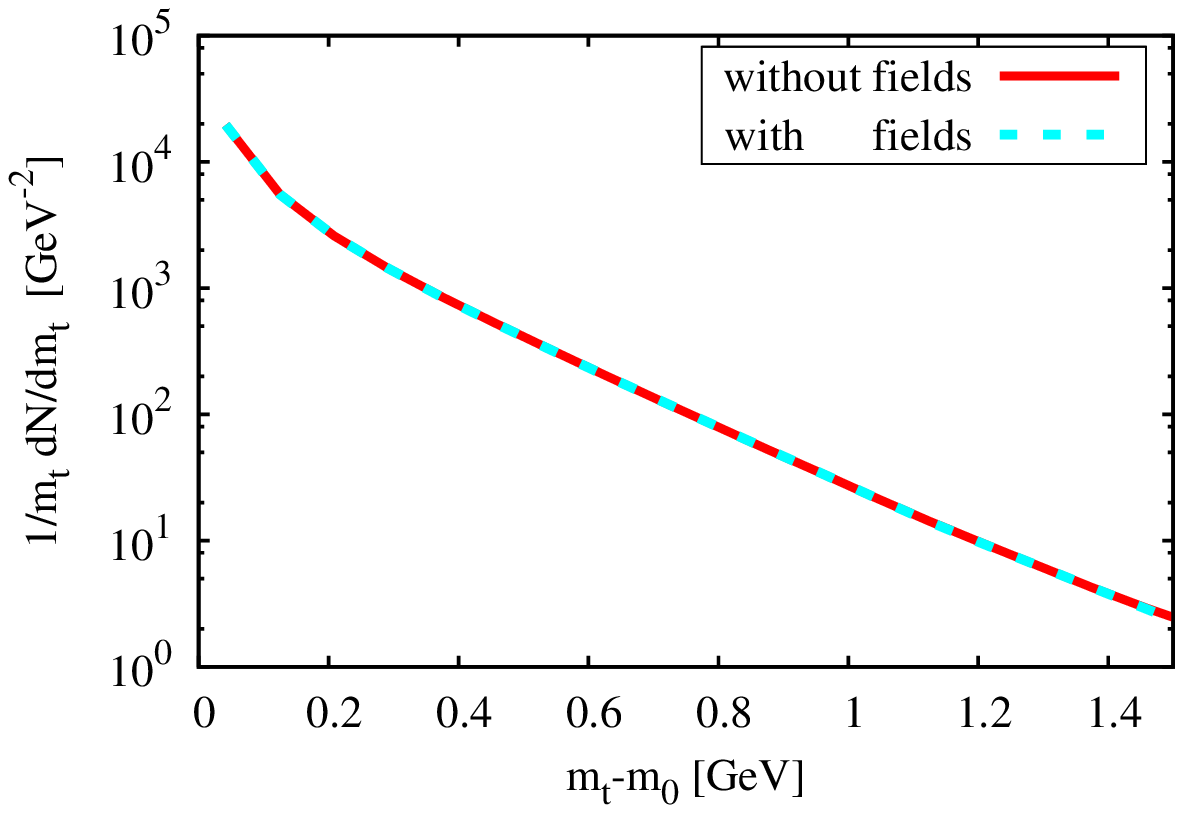}
\hspace*{0.1cm}
\includegraphics[height=4.5truecm,width=6.0truecm] {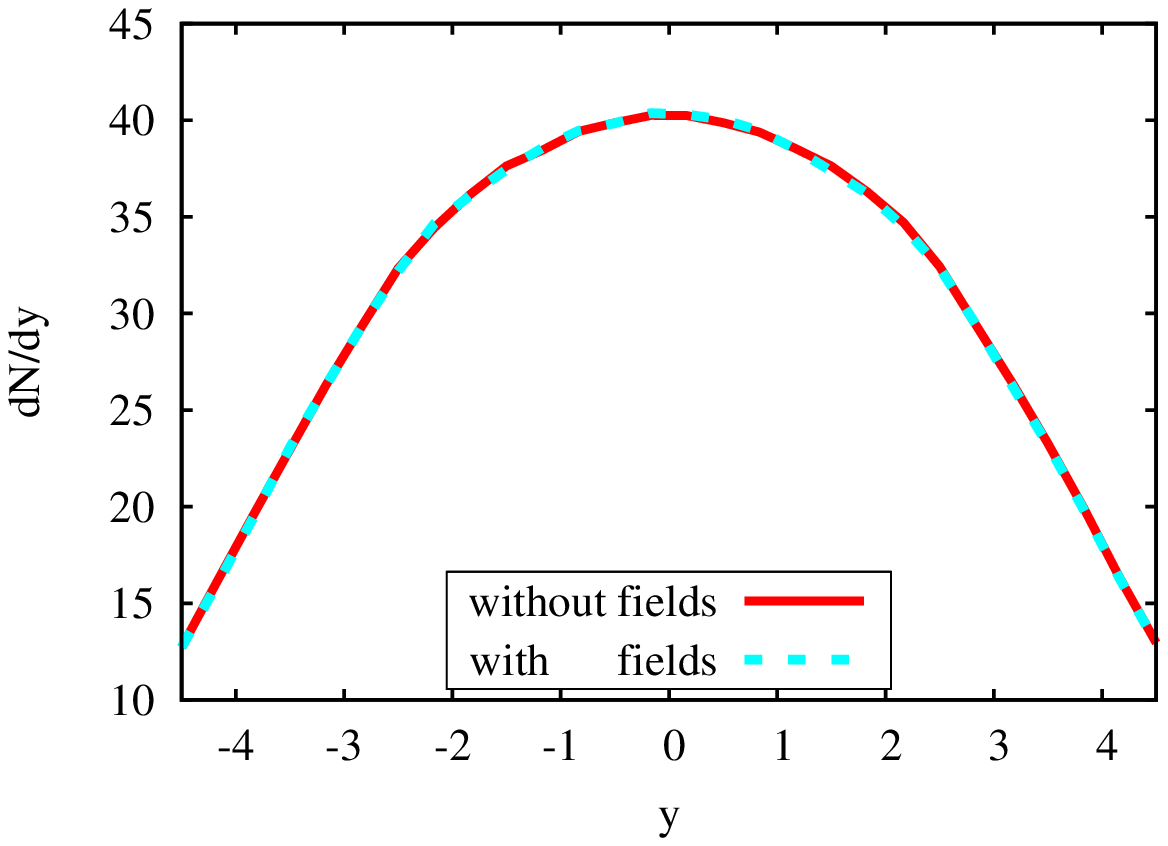}
 \caption{ Transverse mass and rapidity distributions of charged pions
produced in Au+Au ($\sqrt{s_{NN}}=200$ GeV) collisions at $b=10$ fm. 
The results calculated with and without electromagnetic field are plotted 
by the dotted and solid lines, resp.
 }
\label{MtY}
\end{figure}

The HSD results for the versions without  and with electromagnetic field
are presented in Fig.~\ref{MtY}. With a high degree of accuracy,
we see no difference between these
two versions in the transverse mass $m_t$ and rapidity $y$.
\begin{figure}
\centering{
\includegraphics[height=4.5truecm] {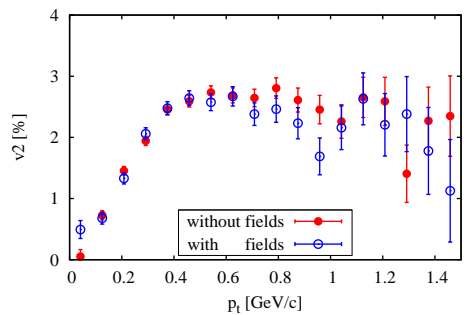}}
\caption{The transverse momentum dependence of the elliptic flow
for Au+Au ($\sqrt{s_{NN}}=$200 GeV) collisions at $b=$10 fm.
 }
\label{v2pt}
\end{figure}
In Fig.~\ref{v2pt} the transverse momentum dependence of the elliptic flow
of charged pions is compared for two versions (with and without field) of the
HSD model. We do not observe any significant difference between the two cases.
Slight differences are
seen in the range of $p_{t}\sim 1$ GeV/c but certainly it can not be 
considered as significant. 
Note that generally the HSD model underestimates the elliptic flow,
but an inclusion of partonic degrees of freedom within the PHSD approach allows
to describe perfectly well the $p_t$ dependence of $v_2$ at the top RHIC
energy~\cite{PHSD}.
\begin{figure}[thb]
\centering{
\includegraphics[width=5.5truecm] {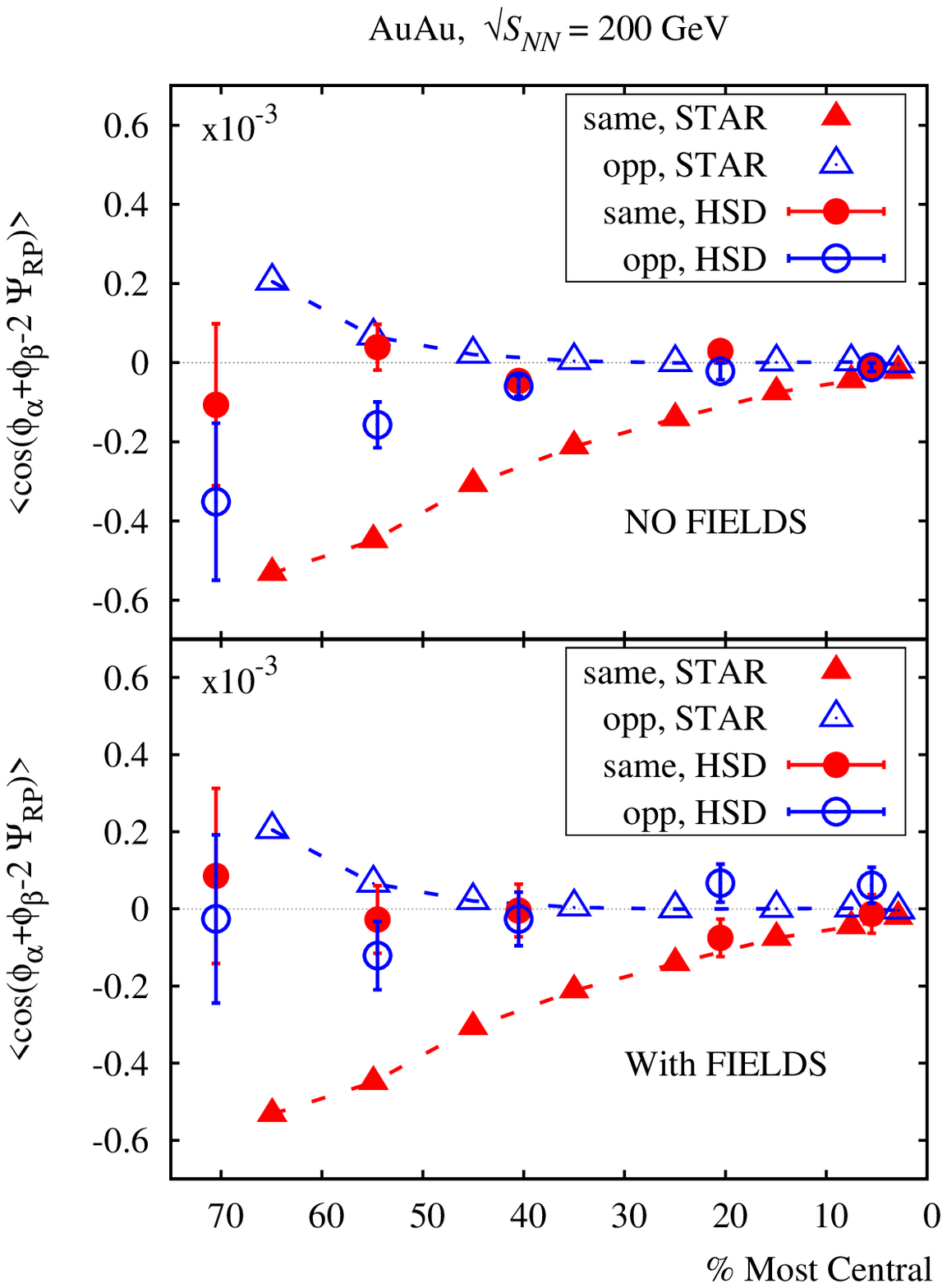} } \hspace*{-0.5cm}
\includegraphics[width=7.0truecm,height=7.7truecm] {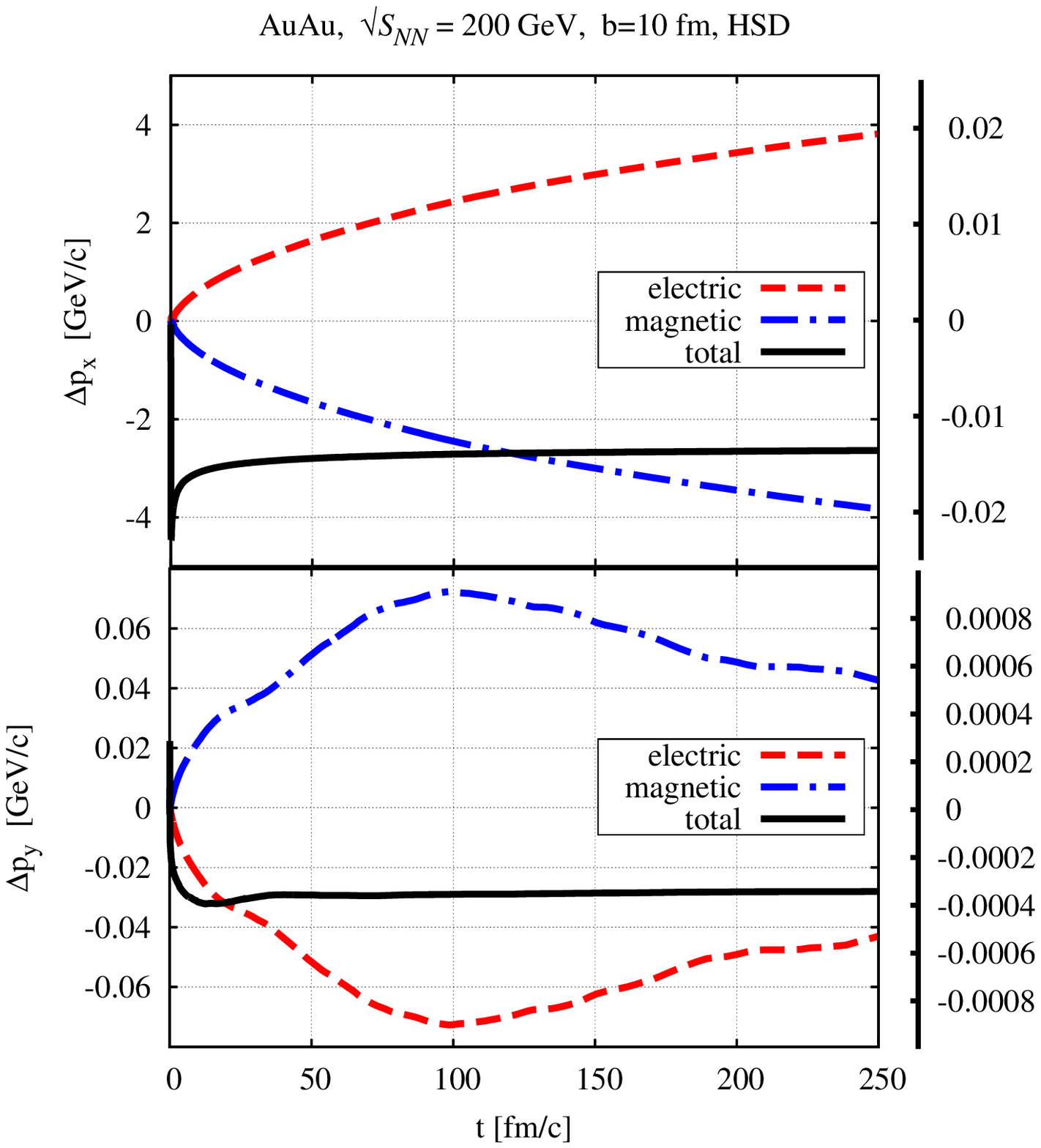}
\caption{ Left: Azimuthal correlation in the transverse plane vs
centrality for like and unlike charged pions from Au+Au
($\sqrt{s_{NN}}=200$ GeV) collisions. The experimental points -
connected by lines - are taken from~\cite{Vo09}. Right: Time
dependence of the momentum increment of mesons created in Au+Au
($\sqrt{s}=200$ GeV) collisions with the impact parameter $b=10$ fm within HSD.
Numbers along the right axis show the total transverse momentum increment 
$\Delta p$ due to electric and magnetic fields. }
\label{C2}
\end{figure}
The correlator (\ref{cos}) is calculated on the event-by-event basis.
The experimental data from the STAR Collaboration~\cite{Vo09} and the
results of HSD calculations are presented in the left panel of Fig.~\ref{C2}.
The experimental acceptance $|\eta |<1$ and $0.15<p_t<2$ GeV were also
incorporated in theoretical calculations.  Note that  the theoretical
reaction plane is fixed exactly by the initial conditions and therefore is not
defined by a correlation with a third charged particle as in the
experiment~\cite{Vo09}.  
As is seen, the HSD model shows no charge separation effect. 
The reason of that is explained in the right panel of Fig.~\ref{C2}.
The average momentum increment $\Delta p$ due to electric and magnetic fields 
is almost completely compensated in every component.

\section{Conclusions}
The model energy dependence of electric  like-charge pairs can be
reconciled with experiment~\cite{Vo09} by a detailed treatment of
the lifetime taking into account both the time of being in a
strong magnetic field and time evolution of the energy density in
the QGP phase. For the chosen parameters we are able to describe
RHIC data for Au+Au collisions on electric charge separation at two
available energies. We predict that the effect will be much
smaller at the LHC energy and will sharply disappear near the top
energy of SPS. 
Coming experiments at the Large Hadron Collider~\cite{ALICE11} 
and that of the planned Beam Energy Scan program  at RHIC \cite{BES11}
are of great interest since they will allow one to test the CME
scenario and to infer the critical magnetic field $eB_{crit}$
governing the spontaneous local {\cal CP} violation.

We have extended the hadron string dynamics model for describing the
formation of the retarded electromagnetic field, its evolution
during a nuclear collision and the effect of this field on
particle propagation. The case of the Au+Au collision at
$\sqrt{s_{NN}}=200$ GeV for $b=10$ fm is considered in great detail.  
It is shown that the most intensive magnetic field oriented
perpendicularly to the reaction plane is formed during the time
when the Lorentz-contracted nuclei are passing through each other,
$t\lsim 0.2$ fm/c. The maximal strength of the magnetic field here
attains very high values, $eB_y/m_\pi^2\sim 5$. 
However, due to the compensation effect, the electromagnetic field does 
not influence on observables and, in particular, on the asymmetry of  
charged mesons with respect to the reaction plane.

\section*{Acknowledgments}
We thank our collaborators E. Bratkovskaya, W. Cassing,
S. Voloshin and V. Konchakovski for useful discussion and remarks.
This work was supported by the LOEWE Center HIC for FAIR.
V.V. is supported in part by the Russian Fund for Basic Research
under grant No. 11-02-01538-a.

\end{document}